\documentstyle[12pt,psfig]{article}
\begin{document}
\begin{titlepage}
\centerline{{\bf ELECTRONIC BAND STRUCTURE IN A PERIODIC}}
\centerline{{\bf MAGNETIC FIELD} \footnote{To appear in {\it Phys.\ Rev.\
B\/} '96} }
\centerline{Andrey Krakovsky \footnote{E-mail: krkvskya@acf2.nyu.edu}}
\centerline{\it Department of Physics, New York University, 
New York, New York 10003}
\centerline{August 18 1995}
\begin{abstract}
We analyze the energy band structure of a
two-dimensional electron gas in a periodic magnetic field of a
longitudinal
antiferromagnet by considering a simple exactly solvable model.  
Two types of states appear: with a finite and
infinitesimal longitudinal mobility.  Both types of states are present
at a generic Fermi surface.
The system exhibits a transition to an insulating regime with respect to 
the longitudinal current, if the electron density is sufficiently low.\\

\noindent PACS number(s): 75.50.Rr, 75.70.Ak

\end{abstract}
\end{titlepage}
The interest in the magnetoconductance properties of the two-dimensional 
electron gas in spatially
periodic lateral magnetic fields has been further stimulated by the recent
experimental availability of such systems \cite{carmona, ye}.  
In the work of
Carmona et al.\ \cite{carmona} spatial modulation of a magnetic
field was produced by
means of equidistantly located superconducting stripes where 
magnetic vortices were
trapped by impurities resulting in periodic inhomogeneity of the
external magnetic field, while in the work of of Ye et al.\ \cite{ye}
it was produced by deposition of ferromagnetic microstructures on
top of the high-mobility two-dimensional (2D) electron gas.  
Vast theoretical efforts on 2D electron gas in an inhomogeneous external
magnetic field range from the theory of 
momentum-dependent tunneling through a magnetic
barrier \cite{matulis} to properties of electronic states and transoprt
in a weakly spatially modulated magnetic field [4--7].
In this short
paper we will be concerned with the one-electron energy band structure
of the 2D electron gas under a periodic lateral magnetic field of an
antiferromagnet, which is a limiting case of a strong periodic
modulation.  We will show that two types of states appear: with a finite
and infinitesimal longitudinal mobility.  Both types of states are
present at a generic Fermi surface.  The system  exhibits a
transition to an insulating regime with respect to the longitudinal
current if the electron density is sufficiently low.

The effect of a uniform magnetic field on energy bands produced by the
periodic (electric) potential is well known \cite{hofstadter}.  
The impact of the slightly inhomogeneous magnetic field on the Landau
levels of a free electron was considered by M\"{u}ller \cite{muller}.  He
showed that the energy bands exhibit a pronounced asymmetry in the
lateral direction.  For
a spatially modulated magnetic field a common
theoretical model \cite{peeters}
employs magnetic field perpendicular to the plane of the
two-dimensional electron gas which has a ``carrier" field $B_0$ with a
periodic modulation on top of it:
\begin{equation}
	{\bf B}=(B_0+B \cos K y){\bf \hat{z}}\,.
	\label{eq:period}
\end{equation}
In the work of Peeters and Vasilopoulos \cite{peeters} the effect 
of a periodic electric 
and weakly modulated magnetic field ($B \ll B_0$) was considered.  They
showed that the broadening of the Landau levels is roughly proportional
to the modulation amplitude $B$.  The ``Hofstadter-like" spectrum was
obtained by Wu and Ulloa \cite{wu}, and collective excitations were
analyzed in \cite{wu2} by the same authors.

In this work we will deal with an electron gas confined to a plane in
a perpendicular periodic magnetic field without a ``carrier" field.  
In other words, in (\ref{eq:period}) we
take $B_0=0$.  This corresponds to the extreme case of the other limit,
$B_0 \ll B_m$.  Such a periodic field will create an energy band
structure of its own.  We see this type of arrangement experimentally
realizable by bringing a two-dimensional 
electron gas in close contact with a mesoscopic longitudinal antiferromagnetic
(sandwich) structure, without an external magnetic field.
When dealing with the one-electron spectrum it is useful to have
some exactly solvable models (potentials), as they elucidate the whole 
structure of the energy bands \cite{slater}.  Below we show that the
energy bands can be obtained exactly in a simple way for a reasonably 
idealized periodic magnetic field.  We present full band picture for
both electron with spin and spinless, and discuss the topology of the
Fermi surface.

The Hamiltonian for a free spinless electron in a magnetic field is:
\begin{equation}
	\hat{H} = \frac{1}{2 m} \left( {\bf p} + \frac{e}{c} {\bf A}
	\right)^2
\end{equation}
In our case the electron is confined to a plane, and a periodic magnetic
field of lateral antiferromagnet is superimposed.  The magnetic field
can be modeled as (see Fig.~1)
\begin{equation}
	{\bf B} = B a \sum_{n=0; \, \pm 1 ; \, ...} 
	 \left[ \delta(y + c + a n) - \delta(y+ a n)
	\right] {\bf \hat{z}}\,,
	\label{eq:field}
\end{equation}
For such a magnetic field the vector potential takes a form
\begin{equation}
	{\bf A} = \big( - B a \epsilon (y), 0, 0 \big)\,,
	\label{eq:gauge}
\end{equation}
where
\begin{eqnarray}
	\epsilon(y) &=& \epsilon(y + a n)\,, \\
	\epsilon(y) &=& \left\{
			\begin{array}{rl}
			-1/2 & 0<y<c \,  \\
			1/2 & c<y<a \,.
			\end{array}
			\right.  
\end{eqnarray}
We proceed with solution of Eqs.~(2)--(6) in a standard way.  We look
for the solution in a form
\begin{equation}
	\psi (x, y) = e^{i k_x x} \chi (y)\,.
\end{equation}
Thus, in the gauge (\ref{eq:gauge}) the solution is a plane wave in the
$x$-direction.  For the $y$-dependent part of the wave function $\chi (y)$
we arrive at
\begin{equation}
	\left[ - \frac{1}{2} \frac{d^2}{dy^2} + k_x \gamma \epsilon (y) \right] 
	\chi (y) = \left(E - \frac{k_x^2}{2} \right)\chi (y)\,.
	\label{eq:kronig}
\end{equation}
Here atomic units are adopted; $E$ is the energy up to an unimportant constant;
$\gamma = a / a_B^2$, the dimensionless ``magnetic length" is given by $a_B =
\sqrt{\hbar c/ e B}/a_0$, $a_0$ is the Bohr radius.  The Eq.~(\ref{eq:kronig})
 is precisely the Schr\"{o}dinger equation for 
the Kronig-Penney model,
and can be easily solved exactly.  The resulting dispersion relation is given by
\begin{equation}
	\cos k_y a = \frac{\beta^2 - \alpha^2}{ 2 \alpha \beta} \sinh
\beta b \sin \alpha c + \cosh \beta b \cos \alpha c \,,
	\label{disp.nospin}
\end{equation}
where 
\begin{eqnarray}
	\alpha &=& \sqrt{k_x \gamma + 2 \left(E - k_x^2/2 \right)}\,;
	\label{eq:alpha} \\
	\beta &=& \sqrt{k_x \gamma - 2 \left(E -  k_x^2/2 \right)}\,;
	\label{eq:beta}
\end{eqnarray}
$k_y$ is the quasimomentum in the longitudinal direction.
The band structure for $B=0.1$T, $a=1\mu$m, $c=a/3$, $b=2a/3$ 
is presented on Figs.~2 and 3.  It
is compressed in the (longitudinal) $y$-direction.  
As was pointed out in \cite{muller}, the pronounced asymmetry along the
$x$-direction is the signature of the energy spectrum in the
inhomogeneuos magnetic field.
In the upper quarter
of Fig.~2 is the region where ``broad" bands are formed.  These bands have
a finite width in the $y$-direction, and the particles occupying these
states will have a finite mobility in the longitudinal direction.  The
other set of ``narrow" 
bands occupies the left and right  quarters of Fig.~2.  From
the point of view of the Kronig-Penney model (\ref{eq:kronig}) they
correspond to the valence bands of the periodic potential.  These bands
are infinitesimally narrow in the $y$-direction, and electrons
populating them would have a vanishingly small longitudinal
mobility.  Of course, in the transverse direction states in 
both types of bands would have some
finite mobility.  Fig.~3 represents the same band structure on a
bigger scale.  The part of the spectrum shown on the Fig.~2 corresponds
to the area inside the box of Fig.~3.  Dark areas of Fig.~3 represent 
regions of ``broad" bands, while the parabolas represent ``narrow"
bands.  In order not to overcomplicate the picture we show only every 
fourth of the latter.  As we will see below, the peculiarity of the
energy spectrum in a
periodic magnetic field will appear in the fact that at the Fermi surface both
types of states will appear.

In this model the problem of electrons with spin is equally easy to
treat.  This amounts to simply adding the spin-dependent term to the
left-hand side of Eq.~(\ref{eq:kronig}):
\begin{equation}
	\left[ - \frac{1}{2} \frac{d^2}{dy^2} + k_x \gamma \epsilon (y) 
	+ \frac{\gamma}{2} \sum_n \big( \delta(y+c+an)-\delta(y+an)\big)\right] 
	\chi (y) = \left(E - \frac{k_x^2}{2} \right)\chi (y)\,;
	\label{eq:kronig.sp}
\end{equation}
which results in a slightly modified dispersion relation:
\begin{eqnarray}
	\cos k_y a &=& \frac{\beta^2 + \gamma^2 - \alpha^2}{2 \alpha
	\beta} \sinh \beta b \sin \alpha a + 
	\frac{\gamma}{\alpha} \cosh \beta b \sin \alpha c \nonumber \\
	& & \mbox{} +
	\frac{\gamma}{\beta} \sinh \beta b \cos \alpha c +
	\cosh \beta b \cos \alpha c 
\end{eqnarray}
with the same $\alpha$ and $\beta$ as in Eqs.~(\ref{eq:alpha}), (\ref{eq:beta}).
The detailed band structure for the same magnetic field as before
is shown on Fig.~4.  The main structure of
the whole spectrum is still represented by the Fig.~3.  As 
compared to the spinless problem, the ``broad" bands are 
characterized by wider gaps in the
density of states, while ``narrow" bands only slightly change their
locations.
As we fill the spin ``up" and ``down" states up to the Fermi level, the
ground state exhibiting transverse oscillatory spin oscillations in the
spirit of the ones discussed by Chudnovsky \cite{chudnov} may result.

Possible Fermi surface 
corresponding to                                         
cutting the energy manifold at $E_F=4 \cdot 10^{-6}$ a.u.\ is presented on
Fig.~5.  We show only a part of
the first Brilloin zone.
Shaded areas are populated by electrons.
Curved lines of the Fermi surface (in the middle)
correspond to states in the ``broad" bands.  As discussed
above, these states have a finite mobility in both directions and
will always contribute to the conductivity of the sample.  The
vertical lines on the extreme right and left 
correspond to the sections of ``narrow" bands which are flat in the
longitudinal direction.  Thus, these states will not contribute to the
longitudinal current, while always contributing to the transverse one.
If the electron gas is dilute enough so that the Fermi level drops below
the $E=0$ level (see Fig.~3), the sample will not conduct in the
longitudinal direction at all as all the states at the Fermi surface 
will have a vanishing mobility in the $y$-direction.  It is easy to
estimate the electron density for transition to an insulating regime by
counting the states with $E <0$.  In our range of parameters we can
totally neglect the width of the ``narrow" bands.
The transition density is given by
\begin{equation}
	n=\frac{1}{2 \pi a} \left( \sum_n \sqrt{\gamma^2 - (\pi + 2 \pi
n)^2/c^2 } + \sum_n \sqrt{\gamma^2 - (\pi + 2 \pi n)^2/b^2} \right)
	\label{eq:transit}
\end{equation}
(all quantities are in atomic units).  The summations are over all $n=0,1
...n_{\rm max}$ for which the radicals remain positive.  For our values
of $b$, $c$, and $B$, $n\approx 10^{11}{\rm cm}^{-2}$.  
If any of the summations turns out to be restricted to the first 
``narrow" band, one has to account for the bandwidth in the
$y$-direction as well.

In a realistic experimental situation many-body effects will be present.
The simplest of them is screening.  Screening will ``smear" the
effective single-particle potential, which may result in suppressing
smaller gaps predicted in the calculation.  However, these effects do
not change the overall structure of the spectrum.

In conclusion, we have presented a simple exactly solvable model of the
electronic band structure in a spatially periodic magnetic field.  
All the conclusions derived from the model are not restricted to this
particular model, but illustrate the general structure of the 
energy bands of the two-dimensional electron gas in a periodic lateral
magnetic field.
This
type of system can be realized by imposing magnetic field of a longitudinal
antiferromagnet on a high-mobility two-dimensional electron gas.  The
band structure exhibits a pronounced asymmetry in the lateral direction
and consists of the two types of bands.  The states in ``broad" bands
will have a finite mobility in the longitudinal direction while the
longitudinal mobility of the electrons occupying states in ``narrow"
bands is infinitesimally small.  A generic Fermi surface will contain
both types of states.  If the electron density is sufficiently low, 
only the ``narrow" bands will be occupied
resulting in vanishing of the longitudinal conductivity.  The electron
density for transition to the insulating regime has been estimated.
An interesting extension of this work is to account for screening in
such a system.

The author would like to thank Professors A. D. Kent, P. M. Levy, J. K.
Percus, and especially J. L. Birman for useful discussions.  
Partial support of the
Physics Department at New York University is gratefully
acknowledged.

\vfill \eject
\section*{Figure Captions}
FIG.\ 1.  The external periodic magnetic field as modeled by (\ref{eq:field}).\\

\noindent FIG.\ 2.  The band structure for a spinless problem compressed
in the (longitudinal) $y$-direction.  $B=0.1$T, $a=3000$\AA, $c=a/3$,
$b=2a/3$.  The states in the ``broad" bands (upper quarter) have a finite
longitudinal mobility, while the longitudinal mobility of the states in
the ``narrow" bands (left and right quarters) is infinitesimal.\\

\noindent FIG.\ 3.  The band structure on a bigger scale.   Only every
fourth of the ``narrow" bands is shown.  The box in
the middle is magnified for the spinless problem (Fig.~2), and the
problem with spin (Fig.~4).\\

\noindent FIG.\ 4.  Same as in Fig.~2 but for the problem with spin.\\

\noindent FIG.\ 5. Fermi surface for $E_F = 4 \cdot 10^{-6}$ a.\ u. Only
a part of the first Brilloin zone is shown.  Shaded areas are populated
by electrons.  The
curved lines of the Fermi surface
(in the middle) correspond to the states in the ``broad"
bands, while the vertical lines on the extreme right and left represent
the states in the ``narrow" bands.
\vfill\eject
%
\end{document}